\documentstyle[aps,twocolumn,overcite,graphicx,epsf]{revtex}
\begin{document}
\def\kmin{k_{min}}
\def\kmax{k_{max}}
\def\ka{k_a}
\def\kb{k_b}
\def\k1{k_1}
\def\ks{k_*}
\def\ksa{k_*^\a}
\def\ir{{\em inviscid-regularization\/ }}
\def\nd{{\em nonlinear-dispersion\/ }}
\newcommand{\emet}{{\em et al.}}
\def\biblio[#1(#2)]#3#4#5#6#7#8{\bibitem{#3} {\sc #4}, {\it #5}, #6 {\bf #7}, #8 (#2).}
\def\submit[#1(#2)]#3#4#5#6{\bibitem{#3} {\sc #4}, {\it #5}, submitted to #6 (#2).}
\def\preprint[#1(#2)]#3#4#5{\bibitem{#3} {\sc #4}, {\it #5}, preprint (#2).}
\def\book[#1(#2)]#3#4#5#6#7{\bibitem{#3} {\sc #4}, {\it #5}, #6, #7 (#2).}
\def\be{\begin{equation}}
\def\ee{\end{equation}}
\def\bea{\begin{eqnarray}}
\def\eea{\end{eqnarray}}
\def\a{\alpha}
\def\fa{\left(1-\a^2\nabla^2\right)}
\def\fai{\fa^{-1}}
\def\qk{\widehat q_\v{k}}
\def\qsk{\widehat q_k}
\def\qsm{\widehat q_m}
\def\qsn{\widehat q_n}
\newcommand{\Ro}{{R\!o}}
\def\v#1{{\bf #1}}
\def\b#1{{\overline{#1}}}

\wideabs{
\title{Scaling properties of an inviscid mean-motion fluid model}
\author{B.T. Nadiga}
\address{Earth and Environmental Sciences, MS-B296, 
	Los Alamos National Laboratory,	Los Alamos, NM 87545}
\maketitle
\begin{abstract}
An inviscid two-dimensional
fluid model with nonlinear dispersion that
arises simultaneously 
in coarse-grained descriptions of the dynamics of the Euler equation 
and in the description of non-Newtonian fluids of second grade 
is considered. 
The scaling of the equilibrium states
of this model for conserved energy and enstrophy retains the corresponding
scaling for the Euler equations on the large scales 
and at the same time greatly deemphasizes the importance of small scales.
This is the first {\em clear} demonstration of the beneficial effect 
of nonlinear dispersion in the model, and should highlight its utility as a subgrid model in more
realistic situations.
\end{abstract}
}

\section{Introduction}
In 1998, Holm et al.~\cite{HMR}, using
the Euler-Poincar\'e variational formalism, proposed a model  
for the mean motion of ideal incompressible fluids.
In this approach,
the (reduced) Lagrangian, which for the 
incompressible case is the kinetic energy, was modified from that 
for the Euler equation:
$$l=\frac{1}{2}\int|\v{u}|^2\, d\v{x} \;\left(=E\right),$$
to account for fluctuation energy of the velocity field in conjunction
with the introduction of a fluctuation length scale $\alpha$:~\cite{bound}
\be\label{erga}
l=\frac{1}{2}\int\left(|\v{u}|^2 + \a^2 |\nabla \v{u}|^2\right)d\v{x}
\;\left(=E^\a\right).
\ee
The resulting ``$\alpha$-model'' for the Euler equations is
$$\frac{\partial \v{v}}{\partial t} + 
	\v{u}\cdot\nabla\v{v} - 
	\a^2\left(\nabla\v{u}\right)^T\cdot\nabla^2\v{u} = - \nabla p$$
\be\label{alp3d}
\nabla\cdot\v{u}=0,\quad
\v{v}=\fa\v{u}
\ee
where, when $\alpha$ is set to zero, $\v{v}=\v{u}$, and 
the usual Euler equations are recovered.
All other notation is standard.  These equations are envisaged as
modeling the flow of inviscid incompressible fluids at length scales
larger than $\a$. (For proof of
existence and uniqueness of solutions of (\ref{alp3d}),
see Shkoller~\cite{S} and
Cioranescu \& Girault~\cite{CG} (viscous case).)

Rivlin and Ericksen~\cite{RE}, in 1955, derived general constitutive laws of the
differential type for an incompressible fluid, wherein
at the first order, viscous
Newtonian stress results (first grade fluids), while at 
the next order inviscid, non-Newtonian, stress-strain relations appear
(second grade fluids). 
Equations (\ref{alp3d}) are identically the equations governing {\em inviscid}
second-grade fluids, and where now $\a$ is a material property.
Viscous and inviscid second grade 
fluid flows have since been studied from different
viewpoints (e.g., see Dunn \& Fosdick~\cite{DF} and references therein,
and Cioranescu \& Girault~\cite{CG}).
We also note that the variational formulation of (\ref{alp3d})
was already {\em explicitly} noted in Cioranescu \& Ouazar~\cite{CO}.

The new derivation of (\ref{alp3d}) has, however, renewed interest in them and
besides spurring more mathematical work has stimulated computational 
investigations of $\a$-models$^{(\hbox{e.g., }}$~\cite{Chen1,Chen2,NS}$^)$
for the reason that the advection velocity, $\v{u}$, is obtained by a spatial-average
of the advected field $\v{v}$
(inversion of the Helmholtz operator in (\ref{alp3d})). This results in a
modification of the advective nonlinearity, the main nonlinearity
of fluid dynamics, in such a way as to suppress mutual
interactions between scales which are smaller than $\alpha$ (as can be
seen, for example, in the untruncated version of (\ref{trunc}) 
below when $|m|,|n| > 2\pi/\a$).
This modification is purely inviscid, and  we will refer to it simply as
\nd in what follows.
However, with the exception of Nadiga \& Shkoller~\cite{NS},
computational studies of $\alpha$-models have always used
additional viscous terms:
For example, Chen et al.~\cite{Chen1} (1998) examine the 
applicability of a viscous $\a$-model to model turbulent channel flow,
and Chen et al.~\cite{Chen2} (1999) explore the utility of a three-dimensional
viscous $\a$-model in providing a subgrid model for fluid turbulence.
While this is clearly the appropriate direction to pursue in the context of
realistic applications, we think that studying purely inviscid $\a$-models,
although idealized, is important and will complement the study 
of their viscous counterparts.
In Nadiga and Shkoller~\cite{NS}, among other things,
we presented a series of two-dimensional
numerical computations comparing the solutions of Euler
equations, Navier-Stokes equations, and an Euler-$\a$
model, and showed that the Euler-$\a$ model was able to reproduce
the typical enstrophy decay characteristics of the Navier-Stokes 
equations, but in a {\em conservative} setting.
Presently, we address some statistical scaling aspects of the
dynamics of such an Euler-$\alpha$ model
to highlight its {\em inviscid} subgrid-scale modeling features.

To better illustrate the effects of {\em nonlinear-dispersion}, 
the salient feature of 
all $\alpha$-models, it suffices
to consider (\ref{alp3d}) in two-dimensions.  
In that case, it can be rewritten in the vorticity-streamfunction 
formulation as
$$\frac{dq}{dt}=\frac{\partial q}{\partial t}
	+J[\psi,q]=0$$
\be\label{alp2d}
q=\fa\omega, \quad
\nabla^2\psi=\omega,
\ee
where $\psi$ is the streamfunction, $\omega$ is the vorticity, $J$ is the
Jacobian operator so that $J[\psi,q]=-{\partial {\psi} \over \partial y} {\partial
q \over \partial x} + {\partial {\psi} \over \partial x} {\partial q
\over \partial y}$, and again, when $\alpha$ is set to zero, $q=\omega$, and
the usual Euler equations result.
Equation (\ref{alp2d}) can be also be written as
$$\frac{\partial\omega}{\partial t}
	+\left(1-\a^2\nabla^2\right)^{-1}
	J[\psi,\left(1-\a^2\nabla^2\right)\omega]=0,
\quad\nabla^2\psi=\omega,$$
a form that highlights the modification to the $J[\psi,\omega]$ nonlinear term
of the Euler equations.
Parenthetically, we note that
in going to two-dimensions, we lose analogs of three-dimensional processes
like vorticity stretching, and therefore, fail to characterize the effect of
\nd $\,$on such processes. 

The kinetic energy $E^\a$ (denoted by $E$ when $\a=0$), 
as defined in (\ref{erga}), is an obvious constant
of motion in both two and three  dimensions.
However, in two dimensions, unlike in three, the vorticity $q$ 
($\omega$ when $\a=0$) of each 
fluid element is an inviscid constant (see (\ref{alp2d})), 
implying an infinity of conservation laws. In particular,
enstrophy $Z^\a$, defined as 
\be\label{ensta}
Z^\a = \frac{1}{2} \int \left[\fa\omega\right]^2 d\v{x},
\ee
is a second conserved quadratic quantity. 
As before, when $\a=0$, we represent the  conserved enstrophy by $Z$.
(The domain integral of $\v{u}\cdot\mbox{\boldmath$\omega$}$ or helicity
is a quadratic quantity which is conserved in three dimensions, but
which is identically zero in two dimensions.)

The use of equilibrium statistical mechanical theories (for
(\ref{alp2d}) with $\a=0$) to better understand the inviscid
dynamics of two-dimensional flows range from the two-constraint
theory (see Kraichnan and Montgomery~\cite{KM} and references therein)
for finite truncations of the continuous system, to those based on
point vortices (again see Kraichnan and Montgomery~\cite{KM} 
and references therein) and their generalizations
to continuous vorticity fields~\cite{RS,MWC} which
consider the infinity of conserved quantities. 
In this article, we
present the two-constraint theory for (\ref{alp2d}) and verify the
main results of the theory computationally.  
Other than mentioning
that there is already some numerical evidence~\cite{NS} which seems to
suggest that individual solutions of the Euler-$\a$ model
(\ref{alp2d}) may indeed follow predictions made for the behavior of
the ensemble-averaged solutions of the Euler equations by the more
complicated statistical theories we do not consider such theories any more
in this short note.  Also, since one may be tempted to point
to the shortcomings of the two-constraint 
theory for the Euler equations before
considering the utility of such a theory for the Euler-$\a$ model, we
wish to point out that the importance of this work
lies primarily in the comparison of the results for the Euler-$\a$
model to the classical results for the Euler equations.  In so doing,
the effects of {\em nonlinear-dispersion}, and its beneficial
numerical ramifications, are clearly highlighted.  At the risk of
belaboring the point further, we reemphasize that in considering the
simple two constraint theory, we are in no way suggesting that the
behavior of the ensemble averaged solution of the $\alpha$-model
(\ref{alp2d}) (or their slightly viscous counterparts) will follow
this theory in more realistic situations; the limitation of this
theory in predicting large-scale coherent structures in the $\a=0$ 
case is well known~\cite{KM}, and carry over to the nonzero $\a$ case.

Furthermore, from a numerical point of view, 
inviscid computations of (\ref{alp2d}) which conserve two quadratic
invariants are fairly easily realizable and
more commonplace than the more involved multisimplectic schemes which
are required for conserving a larger number of constraints. 
Also, while state of the art schemes of the latter
kind can handle only tens of modes (because of an
$N^3\log N$ scaling of computational work, where, $N$ is the
number of modes~\cite{S2}), there is no such restriction on
schemes of the former kind. 
Examples of schemes which conserve
just the energy and enstrophy invariants  are Fourier-Galerkin
truncations implemented as a fully dealiased pseudospectral spatial
discreteization and the second-order finite difference spatial
discreteization using the Arakawa Jacobian.  While we have done
computations with both these schemes and see no discrepancy between
the results, we consider only the spectral discretization in this article
since the theory presents itself most naturally in this setting.

\section{Two-Constraint Statistical Theory for (3)} 
Let $q_\v{x}$ represent a discretization of $q$ on
a two-dimensional spatial grid, $\v{x}$, with $2N+1$ equispaced points
on each side. Let $\qk$, where $\v{k}$ is the set of all wave-vectors
$k=(k_x,k_y)$ denote the Fourier transform of 
$q_\v{x}$. Although there are $(2N+1)^2$ $k$-space
grid points, since $q_\v{x}$ is real, not all of them are independent
and $\qsk=\widehat q^*_{-k}$. Therefore, there are only half as many
$k$-space grid points, and, $\v{k}$ the set of all $k$ is such that
$$\v{k}\equiv\left\{k=(k_x,k_y),\quad \kmin\leq k_x,k_y\leq \kmax\right\}.$$
However, since each $\qsk$ is a complex number,
there are overall $(2N+1)^2$ degrees of freedom in $\qk$.
Consider the truncation of (\ref{alp2d}) that is closed in $\qk$:
\be\label{trunc}
\frac{d}{dt}\qsk + 
	\sum_{\scriptstyle m+n=k\atop \scriptstyle k,m,n\in\v{k}}
	\qsm\qsn\frac{m\times n}
	{|m|^2\left(1+\a^2|m|^2\right)} = 0.
\ee
Among the infinity of conservations for the continuous system (\ref{alp2d})
previously discussed, conservations (\ref{erga}) and (\ref{ensta}) 
are the only ones which survive for the truncated system (\ref{trunc}), and
may be expressed in terms of $\qsk$ as
\be\label{ergnstk}
E^\a = \frac{1}{2}\sum_{k\in\v{k}}\frac{|\qsk|^2}{|k|^2 (1+\a^2 |k|^2)},
\quad Z^\a = 	\frac{1}{2}\sum_{k\in\v{k}}|\qsk|^2.
\ee
This follows from the {\em detailed} conservation property of energy and
enstrophy wherein each of these quantities is conserved in every triad
interaction.

Considering the dynamics of $\qk$ under (\ref{trunc}),
we work in the $(2N+1)^2$ dimensional phase space.
As a consequence of (\ref{alp2d}) satisfying a {\em detailed} 
Liouville theorem
(see Kraichnan and Montgomery~\cite{KM} and references 
therein), (\ref{trunc}) also satisfies a Liouville theorem and 
the motion of $\qk$ in the truncated phase space
is divergence free~\cite{KM}.
We can, therefore, define a stationary probability density, 
$P$, such that $P \prod_{k\in\v{k}}d\qsk$ 
is the probability of finding the system within the 
($(2N+1)^2$ dimensional) 
phase space volume $\prod_{k\in\v{k}}d\qsk$ centered around $\qk$, and
the ensemble average of any quantity $O$, a function of $\qk$,  as
\be\label{ensbl}
\langle O \rangle = \int O P  \,\prod_{k\in\v{k}}d\qsk.
\ee
Next, a maximization of the information theoretic entropy $s$, 
defined in the usual fashion as
$$s=-\langle\ln P - 1\rangle 
=-\int \left(P \ln P - P\right)\,\prod_{k\in\v{k}}d\qsk,$$
subject to constant ensemble-averaged
energy and enstrophy, $\langle E^\a\rangle$ and $\langle Z^\a\rangle$
respectively, leads to 
\be\label{mostprob}
P = a \exp(-\beta E^\a - \gamma Z^\a).
\ee
Here, $\beta$ (an inverse temperature associated with energy) and 
$\gamma$ (an inverse temperature associated with enstrophy)
are the Lagrange multipliers associated with
the two constraints, and $a$ is determined from 
$$ \int P \prod_{k\in\v{k}}d\qsk = 1.$$
Making use of (\ref{ergnstk}) in (\ref{mostprob}) then leads to
a factorization of the probability density:
\be\label{factor}
P = a\prod_{k\in \v{k}}\exp\left(-|\qsk|^2
	\left(\frac{\beta}{|k|^2\left(1+\a^2|k|^2\right)
	} + \gamma\right)\right).
\ee

The ensemble averaged two-dimensional spectral density is then computed 
using (\ref{ensbl}) and (\ref{factor}) (after noting the expressions
for the moments of a Gaussian) as
$$\langle U^\a(k)\rangle \equiv 
\frac{1}{2}\langle\frac{|\qsk|^2}{|k|^2\left(1+\a^2|k|^2\right)}\rangle
=\frac{1}{4} \frac{1}{\beta + \gamma |k|^2\left(1+\a^2|k|^2\right)}.$$
Since (the isotropic) one-dimensional spectra are 
more convenient for plotting, we define
$$E^\a(|k|)= \sum_{|k|\leq|j|< |k|+1} \langle U^\a(j)\rangle,\quad
\hbox{so that}\;\;E^\a = \sum_{|k|} E^\a(|k|).$$ 
In what follows, we drop the $|\cdot|$ sign on $k$ and to avoid confusion, 
note that while $E^\a$ represents the total conserved energy,
$E^\a(k)$, with a dependence on $k$, represents the corresponding 
one-dimensional spectrum.
The one-dimensional spectrum $E^\a(k)$ is then seen to scale with $k$ as
\be\label{ea}
E^\a(k) \sim \frac{k}{\beta + \gamma k^2\left(1+\a^2k^2\right)},
\ee
with the above scaling being only approximate when the mode spacing is not
small compared to $k$ (as at small $k$).

In (\ref{ea}), since $\a$ is a given length scale, 
once the discretization is fixed, expressions for the
total energy and enstrophy of the given initial conditions provide two
equations to solve for $\beta$ and $\gamma$.
The equilibrium spectral scaling (\ref{ea})  is then seen
to exhibit three regimes
depending on the values of the conserved energy and enstrophy as follows.
If the minimum and maximum wavenumbers of the truncation are
$k_{min}$ and $k_{max}$ respectively,
and if we define a mean wavenumber~\cite{k1a} of the initial conditions as,
$$k_1=\sqrt{\frac{Z^\a}{E^\a}},$$
then, we can identify three regimes depending on the signs of 
$\beta$ and $\gamma$:
\begin{itemize}
\item If the initial conditions are such that the
mean wavenumber $\k1$ is small: $\kmin \le \k1 < \ka$, then 
the temperature corresponding to energy is negative, while that
corresponding to enstrophy is positive:
	$-\gamma \kmin^2(1+\a^2\kmin^2)<\beta < 0$,  $\gamma> 0$;
\item If the mean wavenumber $\k1$ is medium:
 $\ka< \k1 < \kb$, then  both temperatures are positive:
$\beta> 0$,  $\gamma> 0$;
\item If the mean wavenumber $\k1$ is large:
$\kb<\k1\le\kmax$, then the temperature corresponding to energy is positive
while the temperature corresponding to enstrophy is negative:
$\beta> 0$, 
	$-\beta<\gamma\kmax^2(1+\a^2\kmax^2)< 0$.
\end{itemize}
Here, $\ka$, and $\kb$ are constants depending on the filter length
$\a$ and the discretization:
$$\ka^2=\frac{\kmax^2-\kmin^2}{2}
\left[\log\left(\frac{\kmax (1+\a^2\kmin^2)}{\kmin(1+\a^2\kmax^2)}\right)\right]^{-1},$$
$$\kb^2=\frac{\kmax^2+\kmin^2}{2} + 
\a^2\frac{\kmax^4+\kmax^2\kmin^2+\kmin^4}{3}.$$
(In the case of an infinite domain, the first of the
above cases, $\beta < 0$, cannot occur since $\ka=0$.)

Further, we can also compute the
spectrum of the energy conserved by the Euler equation ($E$) under the
dynamics of the Euler-$\a$ model. 
Noting that 
$$E = \frac{1}{2}\sum_{k\in\v{k}}\frac{|\qsk|^2}{|k|^2 
\left(1+\a^2 |k|^2\right)^2},$$
(an extra factor $(1+\a^2 |k|^2)$ in the denominator compared to
the expression for $E^\a$)
and that $E$ is not conserved for $\alpha\ne 0$,
the scaling of its one-dimensional spectrum, denoted simply by 
$E(k)$, may be written as~\cite{Evv}
\be\label{ea0}
E(k) \sim \frac{k}{\left(1+\a^2k^2\right)
\left(\beta + \gamma k^2\left(1+\a^2k^2\right)\right)}.
\ee

\section{Discussion and Computational Verification of Results}
We devote the remainder of the article to a discussion of the
scalings (\ref{ea}) and (\ref{ea0}) and
their computational verification.
First, when $\alpha$ is set to zero in either
(\ref{ea}) or (\ref{ea0}), the classic result of Kraichnan~\cite{KM}
for the Euler equation: 
\be\label{classic}E(k) \sim \frac{k}{\beta + \gamma k^2}\ee
is recovered, with the three regions corresponding to the
different combination of signs for $\beta$ and $\gamma$
now separated by values of
the mean wavenumber $\k1$ corresponding to $\ka$ and $\kb$, where 
$\ka$ and $\kb$ are given by
$$\ka^2=\frac{\kmax^2-\kmin^2}{2}
\left[\log\left(\frac{\kmax}{\kmin}\right)\right]^{-1},\quad
\kb^2=\frac{\kmax^2+\kmin^2}{2}.$$

As has been noted many times now~\cite{KM} for $\a=0$, there is no
discontinuity of any sort in going from one region to the
other among the three regions corresponding to different combinations
of signs of $\beta$ and $\gamma$.
Therefore, for convenience, we first consider, in detail,
the case $\beta> 0$ and $\gamma> 0$, and define
$$\ksa=\frac{1}{\a\sqrt{2}}
	\left(-1+\sqrt{\frac{4\a^2\beta}{\gamma} + 1}\right)^{\frac{1}{2}},$$
$$\ks=\lim_{\a\rightarrow 0}\ksa = \sqrt{\frac{\beta}{\gamma}},$$
and note that
$$\ksa=\ks\left(1+O(\a^2\ks^2)\right).$$
Furthermore, $\ks$ can be shown~\cite{BS} to be of the order of $\k1$.
(Thus, for simplicity in what follows, 
one may use $\k1$, $\k1^\a$, $\ks$, and $\ksa$
interchangably, or represent all of them by $\k1$.)
For the Euler solutions, we have from (\ref{ea}) with $\a=0$,
the large scales and small scales (with respect to $\ks$)
behaving asymptotically as
\bea\label{epelr}
&E(k) \sim k, &\quad \kmin \le k\ll \ks;\nonumber\\
&E(k) \sim  k^{-1},&\quad \ks \ll k\le \kmax,
\eea
implying equipartition of $E$ at large scales and equipartition of $Z$ at 
small scales.  (When $k\ll \ks$, $\gamma k^2\ll\beta$ in (\ref{classic})
and when $k\gg\ks$, $\gamma k^2\gg\beta$ in (\ref{classic}).)
When $\a$ is not zero, however, from (\ref{ea}), one easily sees the analogous 
$E^\a$- and $Z^\a$-equipartition results to be respectively
\bea\label{epalp}
&E^\a(k) \sim k, &\quad \kmin\le k\ll \ksa;\nonumber\\
&E^\a(k) \sim  k^{-3},&\quad \ksa\ll k\le \kmax.
\eea
This implies that \nd in (\ref{alp2d}) acts in such a way as to 
{\bf preserve the
Euler scaling of dynamics at the large scales while
at the same time greatly deemphasizing the importance of small scales}.

Asymptotic scalings arising from (\ref{ea0}), for the nonconserved energy
for $\a\ne0$:
\bea\label{aea0}
&E(k) \sim k, &\quad \kmin\le k\ll \ksa;\nonumber\\
&E(k) \sim  k^{-5},&\quad \ksa\ll k\le \kmax.
\eea
further reinforce this result. 
Finally, we note that for $\a=0$, it is well known~\cite{KM} 
(and easy to see from (\ref{ea}))
that when $k_{max}\rightarrow\infty$,
energy diverges  logarithmically and enstrophy diverges
quadratically.
However, when $\a\ne0$, one can see from (\ref{ea}) that
when $k_{max}\rightarrow\infty$, energy is not divergent
and that the enstrophy $Z^\a$ is quadratically divergent.
Nevertheless, in Nadiga and Shkoller~\cite{NS}, we show
that it is the dynamics of the non-conserved enstrophies
$\frac{1}{2} \int \omega^2 d\v{x}$ and
$\frac{1}{2} \int \left[\fa^{\frac{1}{2}}\omega\right]^2 d\v{x}$
which are actually interesting. While the former does not diverge, the
latter diverges only weakly (logarithmically).

We have carried out a series of computational experiments 
on a doubly periodic two-dimensional domain, wherein an ensemble of
initial conditions were evolved under (\ref{trunc}) for different
values of $\a$ until statistical equilibration was achieved.
Initial conditions, similar to those in Fox and Orsag~\cite{FO},
were obtained by choosing amplitudes for wavenumbers 
in the band $50\leq |k| \leq 51$ (zero elsewhere)
from a zero-mean normal distribution of random numbers. The
variance was scaled in such a way that for the different 
values of $\a$, the conserved energy~\cite{Za} had
the same value: $E^\a=E$. The mean wavenumber $k_1$ for this set of initial 
conditions corresponds to about 50.5, and for the resolution
chosen below, $28.5 \le k_a\le 43.5$, and $k_b\ge 60.1$.
With this setup, besides letting both $\beta$ and $\gamma$ be
positive, we can realize both the energy and enstrophy equipartition
regimes in the same experiment.

A fully dealiased pseudospectral spatial discretization was used 
with $\kmin=1$, and $\kmax=N=85$, and
the nonlinear terms were computed in physical space using 256 grid points in
each direction. A nominally fifth-order, adaptive time stepping,
Runge-Kutta Cash-Karp algorithm~\cite{press} was used for time integration.
Energy was conserved to better than 1 in $10^5$ and enstrophy 
to better than 1 in $10^4$ over the entire duration of the runs.

\begin{figure}
\epsfxsize=\columnwidth
\centerline{\epsfbox{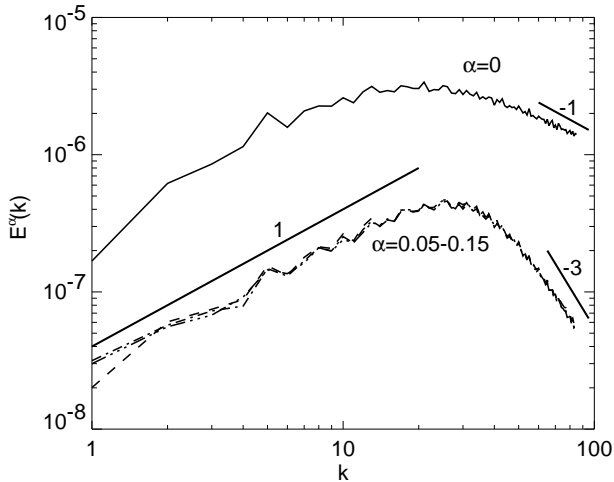}}
\caption{Plot of $E^\a(k)$ vs. $k$ for $\a$=0 (offset by a decade), 
and $\a=0.05$, $0.10$, \& $0.15$. Low-$k$ scaling is identical for zero or
nonzero $\a$, but high-$k$ scaling
is much steeper for $\a\ne 0$. The theoretical asymptotic slopes are drawn
for reference.}
\label{figea}
\end{figure}

In Fig.~\ref{figea}, we plot the (instantaneous) 
spectrum $E^\a(k)$ against $k$, on a log-log
scale for four different values of $\a$: 0, 0.05, 0.10, and 0.15, 
corresponding to
between 0 to $2.4$ percent of the domain size.
The spectrum for the Euler case ($\a=0$) is offset by a decade so as to
not clutter the figure, and slopes of
$1$, $-1$, and $-3$ are drawn for reference.
The scalings, (\ref{epelr}) for $\a=0$, and (\ref{epalp}) for nonzero $\a$,
are clearly verified at both large and small scales,
in Fig.~\ref{figea}, with in fact a cleaner (but identical) scaling
of the large scales for nonzero $\a$.
Furthermore, the three spectra corresponding to the nonzero-$\a$ cases
seem to collapse onto a single curve  in Fig.~\ref{figea}.
This collapse may be explained by 
noting that almost all the energy is contained in the 
low-$k$ modes (denoted below by the set of wavenumbers $\v{k}_{E}$) and almost all
enstrophy is contained in the high-$k$ modes (denoted below
by the set of wavenumbers $\v{k}_{Z}$). This leads to 
a leading order expression for $\beta$  which is independent of
$\a$: $$\beta(\a) \sim \sum_{\v{k}_{E}}k/E,$$
and one for $\gamma$ which is inversely proportional to $\a^2$:
$$\gamma(\a) \sim \sum_{\v{k}_{Z}}k/(\a^2 k_1^2 Z).$$
This in turn implies that the spectra  (\ref{ea})  should be almost
independent of $\alpha$, except for a small intermediate
range of $k$.

\begin{figure}
\epsfxsize=\columnwidth
\centerline{\epsfbox{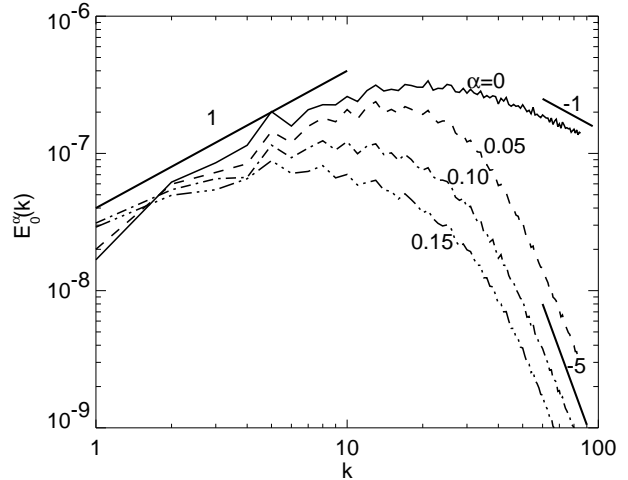}}
\caption{Plot of $E(k)$ vs. $k$ for 
$\a=0$, $0.05$, $0.10$, \& $0.15$. Spectrum falls off much faster ($k^{-5}$)
for $\a\ne 0$ compared to $k^{-1}$ for $\a=0$, but the large-scales are
also increasingly changed with increasing $\a$.}
\label{figea0}
\end{figure}

The above collapse of the  nonzero-$\a$ spectra onto a single 
curve seems to suggest that the actual value of $\a$ is
not very important as long as its value is in a certain range.
This, however, is not true as can be seen from the structure
of the $E(k)$ spectrum plotted in Fig.~\ref{figea0} for the same four
cases. The theoretical scaling for this spectrum is given in (\ref{ea0}) and
the asymptotic scalings are given in (\ref{aea0}).
Although the small scale behavior is as expected,
it is clear that with increasing $\a$, the structure of
the large scales is being substantially modified. (Scaling (\ref{aea0}) at large
scales, can obviously be realized by increasing the number of modes, but that is
not our intent here; we are examining the effect of $\a$ at fixed resolution.)
Therefore, a small value of $\a$ is indicated.
In such a case, the \nd of the $\a$-model 
is highly beneficial at small-scales 
while the large-scale distortion is minimal.
Considering that the minimally resolved length scale in these computations
corresponds to about $0.074$, one may conclude that $\a$ should be of that
order.  That is to say, besides their use in
describing second-grade fluids, $\a$-models in general and (\ref{alp2d}) in particular
should be useful as a subgrid model in under-resolved computations.

\begin{figure}
\epsfxsize=\columnwidth
\centerline{\epsfbox{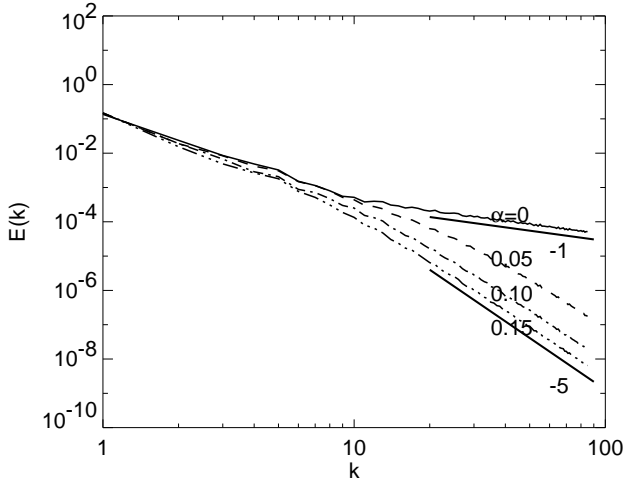}}
\caption{Negative temperature case. Plot of $E(k)$ vs. $k$ for 
$\a=0$, $0.05$, $0.10$, \& $0.15$. Small-scale spectrum falls off 
much faster ($k^{-5}$) for $\a\ne 0$ compared to $k^{-1}$ for $\a=0$.
While the large scale distortion is minimal for $\a$=0.05, it is
appreciable for the other two cases with larger $\a$.}
\label{betan}
\end{figure}

These conclusions are also borne out in numerical experiments
corresponding to the fluid-dynamically 
more interesting case wherein the temperature
associated with energy is negative ($\beta<0$). As mentioned earlier,
such a case is obtained when the initial conditions are chosen
with energy and enstrophy such that $k_1 < k_a$.
(As before, $28.5 \le k_a\le 43.5$ for the different values of
$\a$ for the discretization chosen.)
In such a  case, there is a condensation
of energy on to the low modes of the system~\cite{KM} resulting in 
large scale structures (necessarily coherent).
However, the enstrophy equipartition
scaling of the spectra discussed previously are unchanged.
This gives us an opportunity to better test the
extent of distortion of the low wavenumber (coherent) modes 
due to increasing $\a$.
While various aspects of the negative temperature case for nonzero $\a$ are
considered in Nadiga and Shkoller~\cite{NS}, in the spirit of this
article, we presently consider only the spectral distortion to the
structure of the low wavenumber (coherent) modes.
In Fig.~\ref{betan}, where we plot the spectrum $E(k)$ versus $k$
again for $\a=0$, $0.05$, $0.10$, \& $0.15$, and now where $k_1\approx 2$.
For this case, only one realization (for each value of $\a$)
is considered and the spectrum corresponds to a long time average,
for good measure, taken after the
system has reached statistical equilibrium. 
For this suite of runs, energy was conserved to machine precision while
enstrophy was conserved to about 0.3\% for the entire duration of the
runs considered.
The steep slope of $-5$, for the small scales, for nonzero $\a$ (compared to
a slope of $-1$ for $\a=0$) is again verified and more
importantly, for the case of $\a=0.05$, the low-mode structure (up to 
$k=10$) is almost identical to the case of $\a=0$.
This is clearly not the case for the two other values of the filter
length, $\a$, which are greater than the smallest resolved scale 
of the computation.

I would like to thank Len Margolin, Steve Shkoller, and John Dukowicz
for many interesting discussions.
This work was supported by the Climate Change Prediction Program of DOE at
the Los Alamos National Laboratory.


\begin{references}
\bibitem{k1a}
{We can define $\k1^\a=\sqrt{\frac{Z^\a}{E^\a}}$, and
$\k1=\sqrt{\frac{Z}{E}}$. Then $\k1^\a=\k1\left(1+O(\a^2\k1^2)\right)$.
Since the exact value of $\k1$ is not crucial, to keep things simple, 
we use $\k1$ for $\k1^\a$ also.}
\end{references}
\end{document}